\documentclass[twocolumn,amsmath,amssymb,showpacs,prl,superscriptaddress,aps]{revtex4-1}

\usepackage{epsfig}
\usepackage{color}
\usepackage{graphicx}
\usepackage{bm}
\usepackage{epstopdf}

\begin{document}

\title{Coalescence of Two Impurities in a Trapped One-dimensional Bose Gas}

\author{A.~S. \surname{Dehkharghani}}
\affiliation{Department of Physics and Astronomy, Aarhus University,
DK-8000 Aarhus C, Denmark} 
\author{A.~G. \surname{Volosniev}}
\affiliation{Institut f{\"u}r Kernphysik, Technische Universit{\"a}t Darmstadt, 64289 Darmstadt, Germany}
\author{N.~T. \surname{Zinner}}
\affiliation{Aarhus Institute of Advanced Studies, Aarhus University, DK-8000 Aarhus C, Denmark}
\affiliation{Department of Physics and Astronomy, Aarhus University,
DK-8000 Aarhus C, Denmark}

\date{\today}

\begin{abstract}

We study the ground state of a one-dimensional (1D) trapped Bose gas with two mobile impurity particles. To investigate this set-up, we develop a variational procedure in which the coordinates of the impurity particles are slow-like variables. We validate our method using the exact results obtained for small systems. Then, we discuss energies and pair densities for systems that contain of the order of one hundred atoms. We show that bosonic non-interacting impurities cluster. To explain this clustering, we calculate and discuss induced impurity-impurity potentials in a harmonic trap. Further, we compute the force between static impurities in a ring ({\it {\`a} la} the Casimir force), and contrast the two  effective potentials: the one obtained from the mean-field approximation, and the one due to the one-phonon exchange. Our formalism and findings are important for understanding (beyond the polaron model) the physics of modern 1D cold-atom systems with more than one impurity.

\end{abstract}

\maketitle

\noindent
Doped quasi-one-dimensional cold Bose gases~\cite{kohl2009,catani2012, kuhr2013,knap2017} is a modern platform for studying 1D polarons -- quasiparticles that show up in a medium after renormalizing the impurity parameters~\cite{polaron2000, devreese2009, chevy2010, pietro2014}. The following theoretical model is often used to analyze these systems: A homogeneous and infinite environment with one structureless~\cite{footnote0} mobile impurity particle. Its ground state properties have been understood using quantum Monte Carlo~\cite{parisi2017, grusdt2017}. Other approaches, such as the renormalization group~\cite{volosniev2017, grusdt2017}, perturbation theory~\cite{pastukhov2017}, Bethe ansatz (see~\cite{parisi2017,robinson2016, campbell2017} and references therein), various mean-field (and beyond) approximations~\cite{grusdt2017, volosniev2017, astra2004, sacha2006, bruderer2008,kain2016}, Feynman's variational method~\cite{catani2012,kain2016,grusdt2017} give insight into the limiting cases. This model is widely used, however, care is needed when employing it to describe modern 1D cold-atom set-ups, which (often) have: $i)$ more than one impurity, $ii)$ inhomogeneous densities due to the external confinement, $iii)$~a countable number of bosons (usually from a few to a few hundred~\cite{footnote01}). It is important to add these features to the model because they influence experimental data. For instance, impurity-impurity scattering is blamed for the observed in Ref.~\cite{catani2012} damping of oscillations at vanishing boson-impurity interactions; a trap affects the dynamics at strong boson-impurity interactions~\cite{kamenev2016, grusdt2017, lampo2017}. The effect of the confinement can be qualitatively understood from the polaron model within the local density approximation~\cite{kamenev2016, grusdt2017}. However, to comprehend the role of the impurity-impurity correlations, one must go beyond the polaron model, and investigate systems with at least two impurity particles. Such studies will determine the limits of applicability of the polaron model; moreover, they will guide future research of trapped systems with impurities, and of the corresponding quasiparticles. 

We study the ground state of one such model: A weakly-interacting trapped Bose gas with two impurity particles. We argue that for weak boson-impurity interactions the impurities correlate weakly. The situation changes for strong boson-impurity interactions: The bosonic impurities lower their energy by sharing the same distortion of the gas. Unlike the homogeneous case, this impurity-impurity attraction does not require collective excitations and may be observed even in ideal Bose gases. This attraction can be described by the induced impurity-impurity potential that depends on the center-of-mass and relative coordinates.

\paragraph*{\it Formalism.}

We consider a trapped $(2+N_B)$-body system, which consists of two impurity particles (mass $m_I$) with the coordinates $\{x_i\}$, and  $N_{B}$ bosons (mass $m_B$) with the coordinates $\{y_i\}$.  The corresponding Hamiltonian is
\begin{align}
\begin{split}
\mathcal{H}&=\sum_{i=1}^2 H_{I}(x_i) + \sum_{i=1}^{N_B} H_{B}(y_i) +g_{II}\sum_{j<k}\delta(x_j-x_k)\\
&+g_{IB}\sum_{i=1}^2 \sum_{j=1}^{N_B}\delta(x_i-y_j)+g_{BB}\sum_{j<k}\delta(y_j-y_k)
\label{eq:hamiltonian},
\end{split}
\end{align}
where $H_{I(B)}(x)=-\frac{\hbar^2}{2m_{I(B)}}\frac{\partial^2}{\partial x^2}+V_{I(B)}(x)$ is the one-body Hamiltonian, in which $V_{I(B)}$ is the external trap for the impurities (bosons). The impurity-impurity, boson-impurity and boson-boson interactions are zero-range~\cite{footnote_schur} with strengths $g_{II}$, $g_{IB}$, and $g_{BB}$, respectively. We focus on the $g_{II}=0$ case, which enjoys entirely emergent impurity-impurity correlations. Still, we investigate also the strongly-interacting case ($1/g_{II}=0$) as we consider fermionic impurities. 

Modern experiments often have $N_B\simeq 100$; such mesoscopic set-ups are our primary interest. We develop a formalism built upon our earlier work on a Bose gas with one impurity~\cite{dehkharghani2015a}, adiabatic approximation~\cite{pekar1951}, and the so-called strong coupling approach~\cite{timmermans2006}: We treat the Bose gas as one entity whose density profile quickly adapts to the motion of the impurity, i.e., we assume that $x_1$ and $x_2$ are slow variables and approximate the wave function  $\Psi(x_1,x_2,y_1,\dots,y_{N_B}) $ by
\begin{equation}
\Psi\simeq \phi(x_1,x_2) \Phi(y_1,\dots,y_{N_B}|x_1,x_2),
\label{eq:decomp}
\end{equation}
where $\Phi$ is the ground state of $\mathcal{H}$ with $H_I=0$;
in other words, $\Phi$ describes the ground state of the Bose gas with the impurities fixed at~$x_1$ and~$x_2$. The corresponding energy is~$N_B \epsilon(x_1,x_2)$. Assuming weak boson-boson interactions, we write the function $\Phi$ as the product state: $\Phi=\prod f(y_i|x_1,x_2)$, where $f$ solves a variant of the Gross-Pitaevski equation~\cite{sm}. The decomposition~(\ref{eq:decomp}) might appear similar to the widely used strong coupling approach (cf.~\cite{timmermans2006,blume2006,sacha2006,pelster2016}), in which $\Psi\simeq \chi(y_1)\chi(y_2)\xi(x_1)...\xi(x_{N_B})$, with one-body functions $\chi, \xi$ that minimize the energy.  Note, however, that Eq.~(\ref{eq:decomp}) includes boson-impurity correlations beyond the mean-field approximation, in particular, it entangles the bath with impurities (see also~\cite{chen2017}), which is necessary for accurate results for $g_{IB}\to \infty$. Furthermore, Eq.~(\ref{eq:decomp}) is applicable for bosonic, fermionic, and distinguishable impurities.

Once $f$ and $\epsilon$ are calculated, we obtain the equation for $\phi$ that minimizes the expectation value $E$ of $\mathcal{H}$~\cite{sm}
\begin{equation}
\left(-\frac{\hbar^2}{2m_I}\frac{\partial^2}{\partial x_1^2}-\frac{\hbar^2}{2m_I}\frac{\partial^2}{\partial x_2^2}+V_{eff}\right)\phi=E\phi.
\label{eq:eff_Ham}
\end{equation}
We refer to $E$ as the energy, even though, according to the variational principle, the exact value of the energy is below $E$. The effective potential $V_{eff}$ describes the action of the external trap and the Bose gas on the impurities:
\begin{equation}
\begin{split}
V_{eff}=\sum^2_{i=1} \left(V_I+\frac{\hbar^2 N_B}{2m_I}\left\langle\left(\frac{\partial f(y|x_1,x_2)}{\partial x_i}\right)^2\right\rangle_y\right)+N_B\epsilon.
\label{eq:numericalequation}
\end{split}
\end{equation}
Equations~(\ref{eq:eff_Ham}) and~(\ref{eq:numericalequation}) reduce the original $(2+N_{B})$-body problem to a much simpler two-particle one, in which boson-impurity interactions are hidden in $V_{eff}$. The Schr{\"o}dinger equation~(\ref{eq:eff_Ham}) can be solved using standard numerical routines; we solve it for the ground state~\cite{footnote4} utilizing the finite difference approximation.
  
Note that numerical calculations and analytical analysis~\cite{dehkharghani2015a} showed that such the elimination of the bosonic degrees of freedom leads to accurate results for one trapped impurity (see also~\cite{dehkharghani2015,dehkharghani2016}). Since the ideas put forward in~Ref.~\cite{dehkharghani2015a} do not rely on the number of impurities, as long as $N_B$ is large, it is natural to expect the decomposition~(\ref{eq:decomp}) to work well also for two trapped impurities. 
To verify this anticipation, below we compare to the exact ground state of $\mathcal{H}$, which is computed using the numerical diagonalization method with effective interactions (EEDM) described in detail elsewhere~\cite{dehkharghani2015, lindgren2014}.

%%%%%%%%%%%%%%%%%%%%%%%%%%%%%%%%%%%%%%%%%%%%%

\begin{figure}[t]
\centering
\includegraphics[width=\columnwidth]{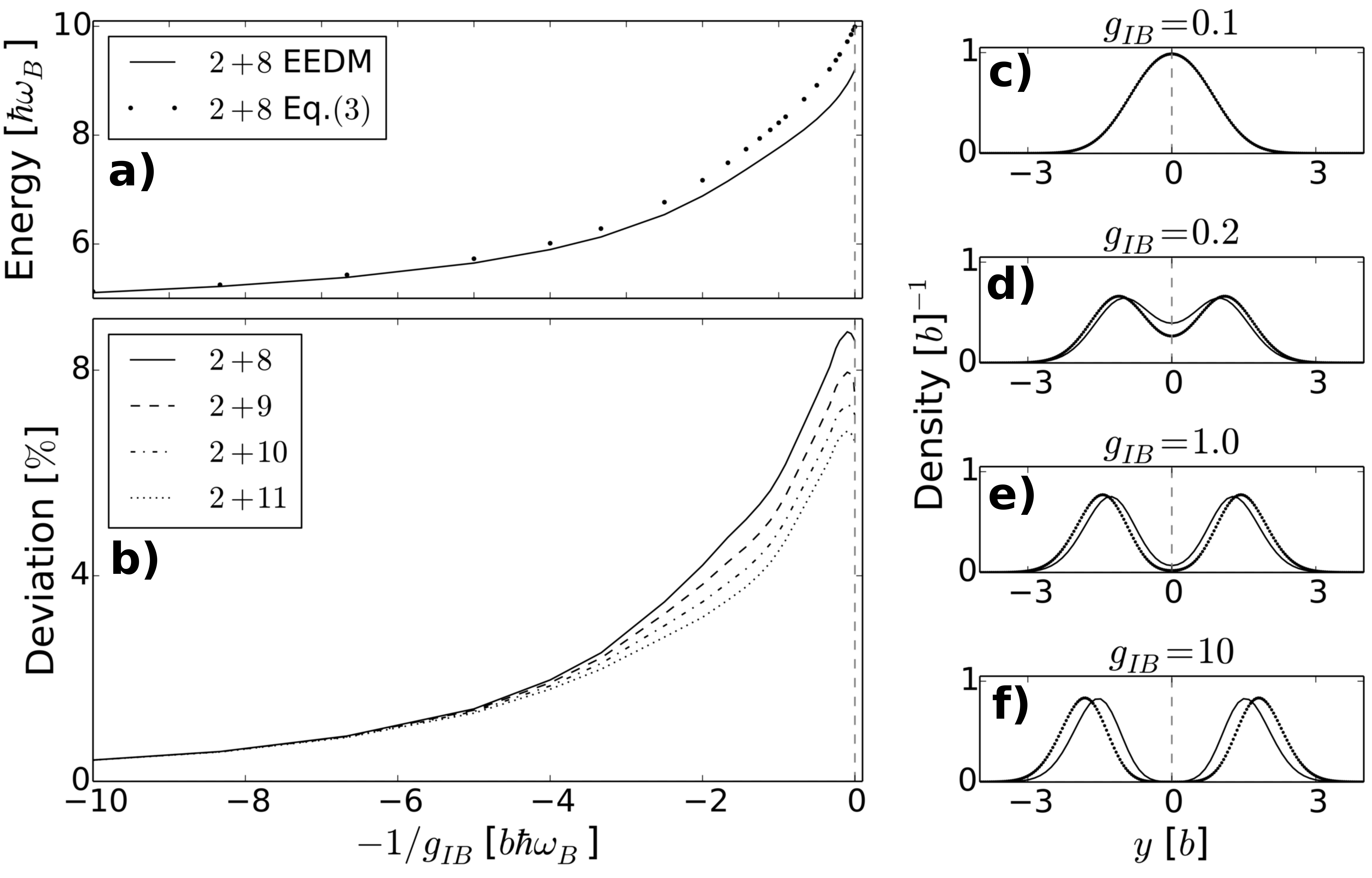}
\caption{ Panels {\bf a)-b)} compare the exact values of the ground state energies with the results of Eq.~(\ref{eq:eff_Ham}).  {\bf a)}: The ground state energy of the $2+8$ system  with $g_{BB}=0$, $m_{I}=m_B$ and $\omega_{I}=\omega_B$. The stars show the exact energy; the dots are obtained from Eq.~(\ref{eq:eff_Ham}). {\bf b)}: The deviation of the energy in Eq.~(\ref{eq:eff_Ham}) from the exact values in \% for a few $2+N_B$ systems with $g_{BB}=0$, $m_{I}=m_B$ and $\omega_{I}=\omega_B$. 
Panels~{\bf c)-f)} compare the impurity densities for the $2+8$ systems (from {\bf a)}) with $g_{IB}=0.1, 0.5, 1$ and $10$ (in units of $b\hbar \omega_B$), respectively. The curves here are calculated using Eq.~(\ref{eq:eff_Ham}), the points depict the exact values. All results are for bosonic impurities.
} 
\label{fig21}
\end{figure}

%%%%%%%%%%%%%%%%%%%%%%%%%%%%%%%%%%%%%%%%%%%%%%%%%%

\paragraph*{\it Particles in harmonic traps.}

%%%%%%%%%%%%%%%%%%%%%%%%%%%%%%%%%%%%%%%%%%%%%%%
\begin{figure}[t]
\centering
\includegraphics[width=\columnwidth]{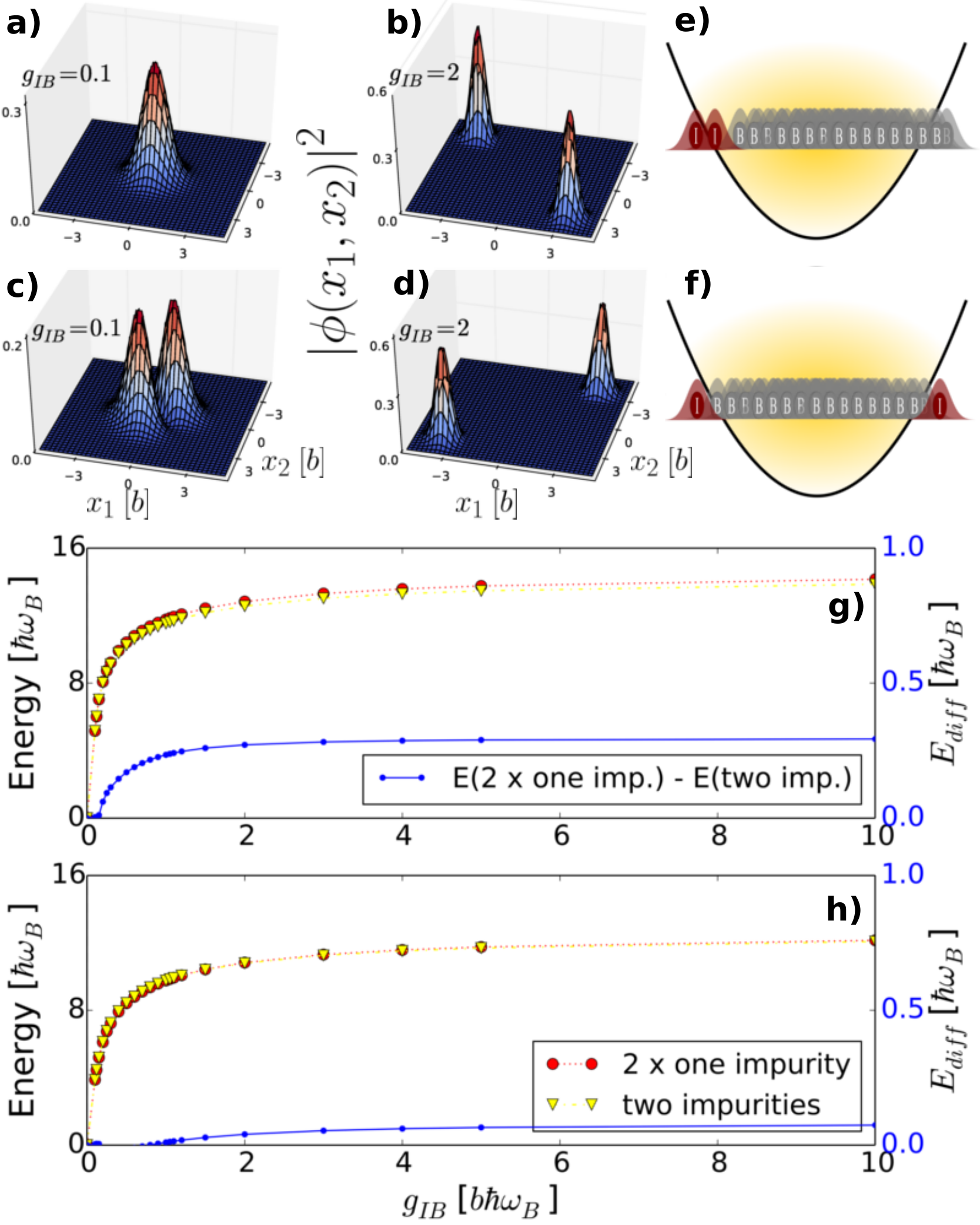}
\caption{Panels {\bf a)} and {\bf b)} show the pair density of the ground state for two bosonic non-interacting impurities placed in an ideal Bose gas (i.e., $g_{BB}=0$) with $N_B=100, m_{I}=m_B$ and $\omega_{I}=\omega_B$  for $g_{IB}=0.1$ and $g_{IB}=2$ (in units of $b\hbar \omega_B$). Panel {\bf g)} is the corresponding energy as a function of the boson-impurity interaction strength $g_{IB}$. The filled circles depict two times the ``one-impurity energy". The triangles are for the ``two-impurities energy", see the text for more details.  The difference $E_{diff}$ is shown with dots (the corresponding $y$-axis is on the right-hand side of the frame). Panels {\bf c)}, {\bf d)} and {\bf h)} describe the densities and energies for fermionic impurities. In panels {\bf e)} and {\bf f)} we sketch the strongly-interacting regimes for bosonic and fermionic impurities, correspondingly.  Note that the lines in $g)$ and $h)$ are to guide the eyes.}
\label{fig1}
\end{figure}
%%%%%%%%%%%%%%%%%%%%%%%%%%%%%%%%%%%%%%%%%%%%%%

Our first application are systems confined by the harmonic traps $V_I(x)=m_{I}\omega_{I}^2x^2/2$ and $V_B(y)=m_{B}\omega_{B}^2y^2/2$. For convenience, we introduce the length unit $b=\sqrt{\hbar/(m_B\omega_B)}$. For small samples the Hamiltonian~(\ref{eq:hamiltonian}) can be diagonalized using EEDM. The comparison of the exact results with those of Eq.~(\ref{eq:eff_Ham}) is shown in Fig.~\ref{fig21}. The agreement between the energies is overall satisfactory. The density profiles are reproduced as well. They show that, to minimize the depletion of the condensate at $g_{IB}\to\infty$, the impurities are driven out of the bosonic cloud.  Similar behavior was observed in the one-impurity case~\cite{dehkharghani2015a, grusdt2017}. The density profile alone does not contain any information regarding the correlations between the impurity particles, thus, below we work with the pair density $n(x_1,x_2)\equiv\int |\Psi|^2 \mathrm{d}y_1 ... \mathrm{d}y_N\simeq |\phi(x_1,x_2)|^2$, which determines the probability to find one impurity particle at $x_1$ if another is placed at $x_2$.

%%%%%%%%%%%%%%%%%%%%%%%%%%%%%%%%%%%%%%%%%%%%%%%%%%%%%%

\begin{figure}[t]
\centering
\includegraphics[scale = 0.22]{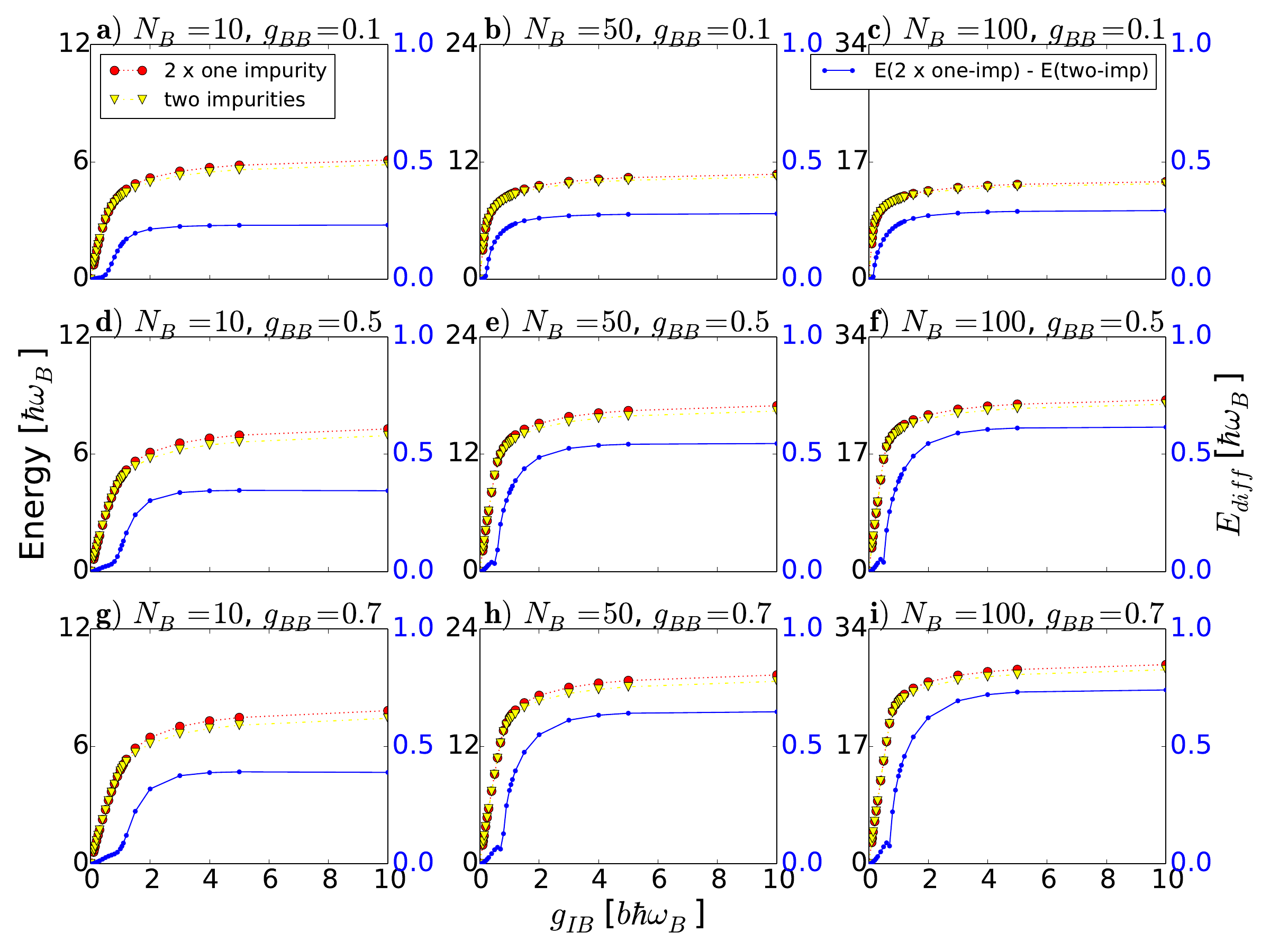}
\caption{The  ground state energy as a function of $g_{IB}$ for a few sets of $N_B$ and $g_{BB}$ (in units of $b \hbar \omega_B$). Here $m_{I}=m_B$ and $\omega_{I}=\omega_B$. The filled circles depict two times the ``one-impurity energy". The triangles  are for the ``two-impurities energy", see the text for details. The difference $E_{diff}$ is shown with dots. The lines are to guide the eyes.}
\label{fig2}
\end{figure}

%%%%%%%%%%%%%%%%%%%%%%%%%%%%%%%%%%%%%%%%%%%%%%%%%%%%%%%%

Having tested the formalism we turn to the $2+100$ system  with $g_{BB}=0$, which is too complex for EEDM.  
We consider bosonic and fermionic impurities~\cite{footnote2}; see Fig.~\ref{fig1}, where two limits are clearly seen. When the interactions are weak a simple picture of two non-interacting particles describes well the system; see Figs.~\ref{fig1}{\bf a)} and {\bf c)}. In this case the Bose gas can be treated simply as an external potential~\cite{footnote1} for the impurities. 
For $g_{IB}\to\infty$ the impurities move to the edge of the trap just as for $N_B=8$; see Figs.~\ref{fig1}{\bf b)} and {\bf d)}. The pair density shows that the bosonic impurities cluster and move to the edge as a whole;  see Fig.~\ref{fig1}{\bf b)}.  This behavior can be explained by an overall attractive impurity-impurity interaction potential (see below) due to  the trap: Impurities in a homogeneous infinite Bose gas do not correlate if $g_{BB}=0$. This interaction, however, is not strong enough to bind two fermionic impurities; see Fig.~\ref{fig1}{\bf d)}.

The impurity-impurity attraction also leaves a footprint in the energy domain.  To unravel it, we calculate the ``two-impurities energy" $\mathcal{E}_2=E(g_{IB})-E(g_{IB}=0)$ and compare it to two times the ``one-impurity energy", which is  $2\mathcal{E}_1=2(\varepsilon(g_{IB})-\varepsilon(g_{IB}=0))$ for bosons \cite{fermions}, where $\varepsilon(g_{IB})$ is the energy of the $1+N_{B}$ system calculated as in~Ref.~\cite{dehkharghani2015a}. If the impurities do not correlate then the energy difference between the latter and the former, $E_{diff}=2\mathcal{E}_1-\mathcal{E}_2$, is zero; $E_{diff}>0$ means a positive ``binding energy", and, hence, effective attraction between the impurities; $E_{diff}<0$ can also happen due to a finite size of our system.  At weak interactions $E_{diff}\simeq 0$, thus, two impurities can be well described as two non-correlated particles; see Figs.~\ref{fig1}{\bf g)} and {\bf h)}. As $g_{IB}$ increases the two bosonic impurities form a ``bound state", whereas two fermionic impurities do not; see Figs.~\ref{fig1}{\bf e)} and {\bf h)}. We mention in passing that there are other systems in which impurities strongly attract each other. For example, materials in which electrons, even in spite of their fermionic nature and Coulomb repulsion, form a new bosonic bound state -- bipolaron, the quasiparticle, which might play a role in the properties of high-T$_c$ superconducting materials (see~\cite{devreese2009} and references therein). Another famous example is a superfluid $^4$He doped with $^3$He atoms~\cite{pines1966, baym1966}. It is tempting to associate our bound states with bipolarons. However, we prefer to leave the discussion of quasiparticles for future studies and focus here on studying the ground state correlations.

The clustering demonstrated in Fig.~\ref{fig1} also happens for other values of $N_B$ and $g_{BB}$~\cite{footnote3}. We illustrate this for bosonic impurities in Fig.~\ref{fig2}; the correlations of fermions seem to be driven mainly by the Pauli exclusion principle and we refrain from showing more on this case. Similarly to Fig.~\ref{fig1}, the curves in Fig.~\ref{fig2} have distinct behaviors at small and large values $g_{IB}$. However, there are some noticeable changes: The energy difference $E_{diff}(g_{IB}\to\infty)$ grows with $g_{BB}$, the transition region from the weakly- to strongly-correlated regimes is shifted. The shift is intuitively clear: $g_{IB}$ should be larger than $g_{BB}$ to push the impurities out of the bosonic cloud. These correlations agree qualitatively with the phase separation of two harmonically-trapped Bose gases~\cite{timmermans1998, malomed2000}. 

The clustering may be detected in cold-atom systems. For example, by changing $g_{IB}$ from zero to some large value: When the system reaches equilibrium, the probability to observe impurities together on one edge of the trap will be larger than in the non-correlated case. The energy $E_{diff}$ defines the temperature window, in which this entanglement of impurities is important. For the considered cases, $E_{diff}$ is of the order of $\hbar \omega_B$ (see Fig.~\ref{fig2}), which for typical values of $\omega_B/(2\pi)=(0.1-1)$ kHz is $\sim (4-40) n \mathrm{K}\times k_B$, here $k_B$ is the Boltzmann constant. If $g_{II}>0$ this number must become smaller; it is zero, if the impurities fermionize, i.e., $1/g_{II}=0$.

%%%%%%%%%%%%%%%%%%%%%%%%%%%%%%%%%%%%%%%%%%%%%%%%%%%%%%

\begin{figure}[t]
\centering
\includegraphics[width=\columnwidth]{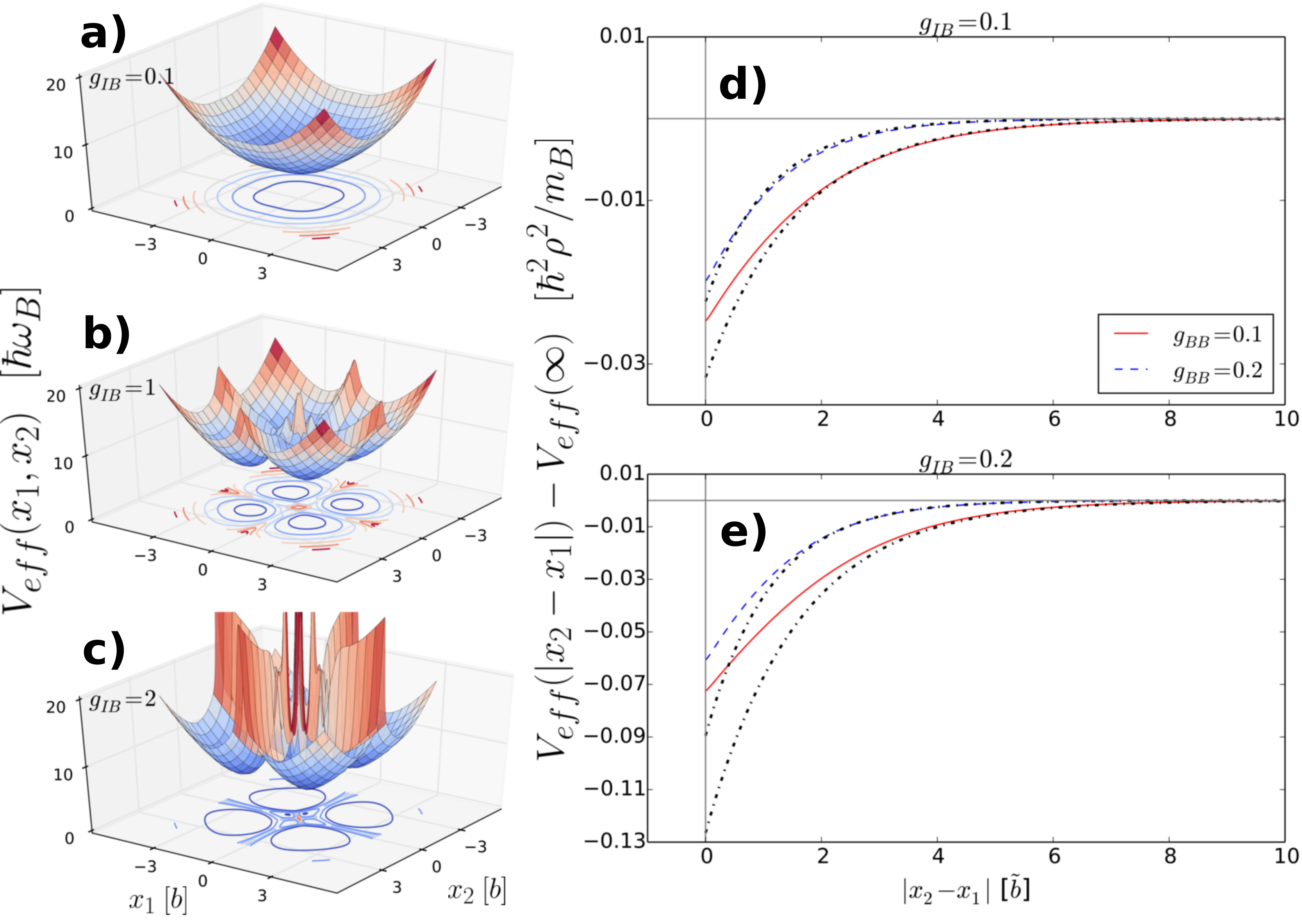}
\caption{Panels {\bf a)-c)} show the effective potential~(\ref{eq:numericalequation}) for the trapped $2+8$ system considered in Fig.~\ref{fig21}. Panels {\bf d)} and {
\bf e)} show the potential for two heavy impurities in a ring in the thermodynamic limit as a function of the relative distance. The dot-dashed (black) curves here show the corresponding Yukawa potential, i.e., 
$-\sqrt{\frac{m_B\rho g_{IB}^4}{\hbar^2 g_{BB}}}\mathrm{exp}\left(-2\sqrt{\frac{m_B g_{BB}\rho}{\hbar^2}}|x_2-x_1|\right)$. The parameters $g_{IB}$ and $g_{BB}$ in {\bf a)-c)} are measured in units $b\hbar\omega_B$; in {\bf d)} and {
\bf e)} -- in units of $\hbar^2\rho/m_B$. }
\label{fig:potential}
\end{figure}

%%%%%%%%%%%%%%%%%%%%%%%%%%%%%%%%%%%%%%%%%%%%%%%%%%%%%%

\paragraph*{\it Effective interaction.}  

 The bunching of bosonic impurities follows directly from the form of the effective potential~$V_{eff}$; see Fig.~\ref{fig:potential}. Weak interactions lead to small modifications to $V_{I}(x_1)+V_{I}(x_2)$. Larger values of $g_{IB}$ drastically affect $V_{eff}$, which now has minima at the edges of the bosonic cloud. The minima at the $x_1=x_2$ line
are the deepest leading to the coalescence of impurities.  
We stress again that $V_{eff}$ incorporates the effects of both the trap and the Bose gas. For small trapped systems, it seems impossible to disentangle these forces (see also~\cite{chen2017}) and obtain an effective impurity-impurity interaction from $V_{eff}$ that depends only on $|x_1-x_2|$, which is expected in a homogeneous case, see, e.g.,~\cite{stoof2000, kamenev2014}.

A thorough study of homogeneous systems is beyond this paper's scope. To connect our research to previous studies, we discuss $V_{eff}$ for two static impurities trapped in a 1D ring-shape trap of length~$L$. This set-up allows us to exclude the trap effects, and to use the existing results on the Casimir-like forces~\cite{recati2005, bruderer2007} for comparison.  We focus on the thermodynamic limit~\cite{footnote02}, i.e., $N_B (L)\to\infty$ and $N_B/L\to\rho$; the length unit now is $\tilde{b}=1/\rho$. 
We calculate the quantity $V_{eff}(|x_2-x_1|)-V_{eff}(\infty)$, which determines the induced impurity-impurity interaction; see Figs.~\ref{fig:potential}{\bf d)-e)}. For comparison, we also plot the Yukawa potential (YP), which is expected~\cite{footnote_diff} in the perturbative regime from the one-phonon exchange (see~\cite{sm,recati2005,naidon2016}). The graphs show that for $g_{IB}\to 0$ the YP describes well the long-range physics and should be slightly corrected at small values of $|x_2-x_1|$. For $x_2=x_1$ the correction can be calculated analytically~\cite{sm}. For large values of $g_{IB}$ the YP fails, e.g., it predicts infinite attraction for $g_{IB}\to\infty$. In this regime our method provides means for calculating emergent impurity-impurity potentials, which are not the YP. These potentials are attractive at large distances within our mean-field approach~\cite{sm}. Note that tails of induced impurity-impurity interactions can in principle have repulsive regions if one includes quantum fluctuations~\cite{kamenev2014}, which are not considered in the present Letter.

The potential $V_{eff}$ is generated by the mean-field transformation of the Bose gas' density, hence, we shall refer to $V_{eff}$ as the mean-field part of the effective force (motivated by the studies of thin films~\cite{zandi2007,biswas2010}).
In principle, $V_{eff}$ can be measured by doing spectroscopy with two tightly trapped atoms~\cite{recati2005, klein2005}.

Within our formalism, effective potentials for mobile impurities look similar to the presented in Figs.~\ref{fig:potential}{\bf d)} and {\bf e)}. They have negative net volume~\cite{footnote_separation}, hence, they support two-body bound states~\cite{landau1977} even for $g_{IB}\to 0$. For $g_{IB}\to 0$, though, the binding energy is small~\cite{footnote_bind_energy} and finite temperature will wipe out any emergent correlations. By increasing $g_{IB}$ the bound state becomes ``deeper" and can be observed. One might describe this limit as the dark-bright soliton (see~\cite{Kevrekidis2016}), which, in the present context, corresponds to a multi-polaron bound state that possesses some properties of a quasiparticle.

\begin{acknowledgments}
This work was supported by the Danish Council for
Independent Research and the DFF Sapere Aude program. 
A. G. V. gratefully acknowledges the support of the Humboldt Foundation. We thank Manuel Valiente, Thomas Pohl, Fabio Cinti and Luis Ardila for enlightening discussions about the Casimir-like force. We thank Jens Braun for useful conversations and comments on the manuscript. 

Note added: See the recent study in Ref.~\cite{reichert2018}
for details
on the contradicting results in the literature described
in the footnote in Ref.~\cite{footnote_diff}.
\end{acknowledgments}

\appendix

\widetext

\section{Technical details}\label{app}

\subsection{Derivation of Eqs.~(3) and (4)}
\label{ch:polarons:sec:doublequantumimpurity}
\begin{figure}[t]
\centering
\includegraphics[width=\columnwidth]{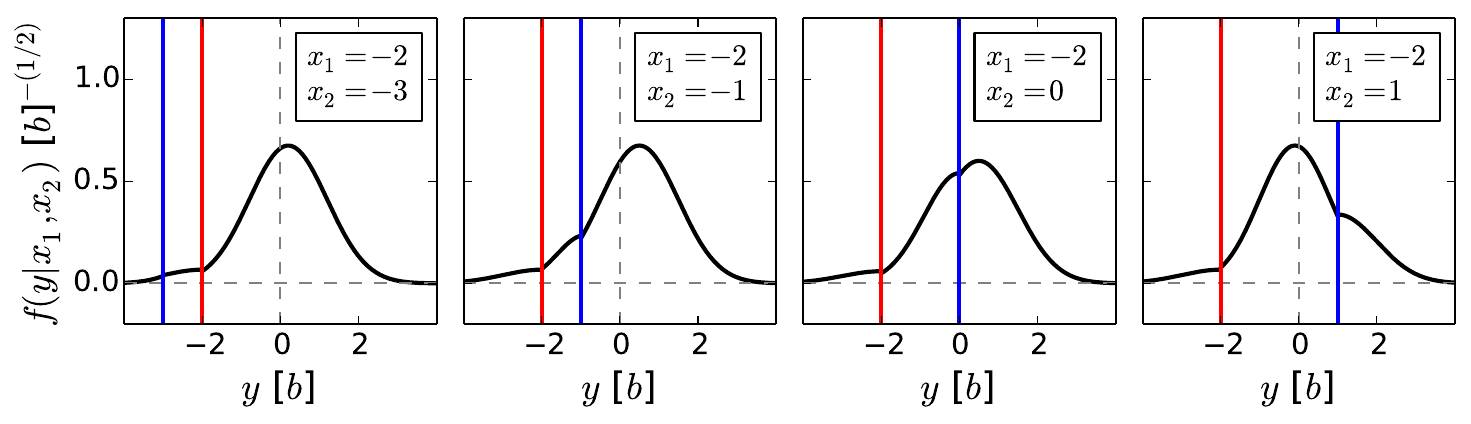}
\caption{A sketch of the function $f(y|x_1,x_2)$ for $g_{BB}=0$ in the harmonic oscillator with $m_I=m_B$ and $\omega_I=\omega_B$. The lines show the positions the impurities.}
\label{polaronsketch2+N}
\end{figure}

For the system with two impurities fixed at the positions $x_1$ and $x_2$, a complete basis is given by the set of functions $\Phi_j(y_1,...,y_{N_{B}}|x_1,x_2)$ that solve the equation:
\begin{equation}
\left(\sum_{i=1}^{N_B} H_{B}(y_i)+g_{IB}\sum_{i=1}^2 \sum_{j=1}^{N_B}\delta(x_i-y_j)+g_{BB}\sum_{i<j} \delta(y_i-y_j)\right)\Phi_j=E_j(x_1, x_2)\Phi_j.
\label{eq:app1:1}
\end{equation}
By assumption, the set of functions $\Phi_j$ is orthonormal in the $y$-space. Using this basis we write the wave function $\Psi$ for the system with two mobile impurities
\begin{equation}
\Psi(x_1,x_2,y_1,...,y_{N_B})=\sum_{j}\phi_j(x_1,x_2)\Phi_j(y_1,...,y_{N_{B}}|x_1,x_2),
\end{equation} 
where the functions $\phi_j(x_1,x_2)$ define the decomposition. We normalize the function $\Psi$ to one, and write the corresponding expectation value as
\begin{equation}
E=\sum_{i,j} \int \mathrm{d}x_1\mathrm{d}x_2 \phi_j(x_1,x_2) \left(\sum_{l=1}^2 \left (H_{A}(x_l)-P_{ji;l}(x_1,x_2)\frac{\hbar^2}{m_I}\frac{\partial}{\partial x_l} 
 + Q_{ji;l}(x_1,x_2)\right) +E_{j}(x_1,x_2)\delta_{ij}\right)\phi_i(x_1,x_2), 
\end{equation}
where $\delta_{ij}$ is the Kronecker delta, and the functions $P_{ji;l}$ and $Q_{ji;l}$ are defined as
\begin{align}
P_{ji;l}(x_1,x_2)=\int \mathrm{d}y_1 ... \mathrm{d}y_{N_B} \Phi_j(y_1,...,y_{N_B}|x_1,x_2) \frac{\partial}{\partial x_l}\Phi_i(y_1,...,y_{N_B}|x_1,x_2), \\
Q_{ji;l}(x_1,x_2)=-\frac{\hbar^2}{2m_I}\int \mathrm{d}y_1 ... \mathrm{d}y_{N_B} \Phi_j(y_1,...,y_{N_B}|x_1,x_2) \frac{\partial^2}{\partial x_l^2}\Phi_i(y_1,...,y_{N_B}|x_1,x_2).
\end{align}
Note that due to the normalization condition on $\Phi_i$ the coupling $P_{ii;l}$ is zero. In the main text we assume that the boson-boson interaction is either vanishing or small, so that the (quasi) Bose-Einstein condensate plays the main role. Therefore, we use only  the lowest-energy solution of Eq.~(\ref{eq:app1:1}), which yields 
\begin{equation}
E\simeq \int \mathrm{d}x_1\mathrm{d}x_2 \phi(x_1,x_2) \left(\sum_{l=1}^2 \left (H_{A}(x_l) 
 + Q_{00;l}(x_1,x_2)\right) +E_0(x_1,x_2)\right)\phi(x_1,x_2).
\end{equation}
Here we have used the notation of the main text $\phi=\phi_0$.  The value $E$ is a variational upper bound on the energy. It is minimized if $\phi$ is the ground state of the Hamiltonian
\begin{equation}
h= \sum_{l=1}^2 \left[(H_{A}(x_l) 
 + Q_{00;l}(x_1,x_2)\right] +E_0(x_1,x_2).
\end{equation}
To find $E_0$ and $Q_{00;l}$ we write $\Phi=\prod f(y_i|x_1,x_2)$, where $f(y_i|x_1,x_2)$ solves the Gross-Pitaevski equation that corresponds to the Schr{\"o}dinger equation~(\ref{eq:app1:1}) (see the next section). We plot it in Fig.~(\ref{polaronsketch2+N}). The function $E_0$ is then simply $N_{B} \epsilon(x_1,x_2)$, where $\epsilon$ is 
\begin{equation}
\epsilon(x_1,x_2)=\int \mathrm{d}y f(y|x_1,x_2)\left( H_B(y) + g_{IB}\sum_{i=1}^2\delta(x_i-y) + \frac{g_{BB}}{2}(N_{BB}-1) f(y|x_1,x_2)^2\right)f(y|x_1,x_2);
\end{equation}
the coupling $Q_{00;l}$ has the form
\begin{equation}
Q_{00;l}(x_1,x_2)=N_B\frac{\hbar^2}{2 m_I}\int \mathrm{d}y \left(\frac{\partial f(y|x_1,x_2)}{\partial x_l}\right)^2.
\end{equation}

It is worthwhile noting that the predictive power of our approach can be improved if the couplings $P$ and $Q$ are included. For an ideal Bose gas  in a harmonic trap, the function $\Phi_i$ can be found analytically. For a weakly-interacting Bose gas in a trap, the Bogoliubov approximation should be employed to obtain elementary excitations of the Gross-Pitaevski equation. We leave this investigation for future work.

\subsection{Two static impurities in a Bose gas}

\noindent The system with two static impurities within our formalism is described by the Gross-Pitaevski equation  
\begin{equation}
-\frac{1}{2} \frac{\mathrm{d}^2f(y)}{\mathrm{d}y} + g_{BB}(N_{BB}-1)f(y)^3+g_{IB}\left(\delta(x_1-y)+\delta(x_2-y)\right)f(y)=\mu f(y),
\label{app:eq:gpe}
\end{equation}
where $\mu$ is the chemical potential, and for convenience, we set $\hbar=m_B=1$.  For simplicity, in what follows we assume that $x_1=0$ and $x_2=r>0$.

{\it Weakly-interacting case.} Let us start by considering a weakly-interacting case, i.e., $g_{IB}\to 0$, in the thermodynamic limit, i.e., $N_{BB}=\rho L\to \infty$. We assume that $f=(1+F)/\sqrt{L}$, $\mu=\mu_0+\mu_1$, where $\mu_0=\frac{g_{BB}(N_{BB}-1)}{L}$; $F$ and $\mu_0$ solve the following equation
\begin{equation}
-\frac{1}{2} \frac{\mathrm{d}^2 F}{\mathrm{d}y^2} + \mu_0 (3F+3F^2 + F^3)+g_{IB}\left(\delta(x_1-y)+\delta(x_2-y)\right)(1+F)=\mu_1 + \mu_0 F +\mu_1 F.
\end{equation}
Since $g_{IB}$ is small we write $F=F_0+F_1+...$, such that
\begin{align}
-\frac{1}{2} \frac{\mathrm{d}^2 F_0}{\mathrm{d}y^2} + 2 \mu_0 F_0+g_{IB}\left(\delta(x_1-y)+\delta(x_2-y)\right)=\mu_1, \\
-\frac{1}{2} \frac{\mathrm{d}^2 F_1}{\mathrm{d}y^2} + 2 \mu_0 F_1 + 3\mu_0 F_0^2+g_{IB}\left(\delta(x_1-y)+\delta(x_2-y)\right) F_0=\mu_1 F_0, \\
...\qquad .
\end{align}
We truncate this set of equations at $F_2$ and beyond, which will provide the energy in the $g_{IB}^2$ order. The truncated system can be easily solved.
The function $F_0$ is  
\begin{align}
F_0(y) = \frac{\mu_1}{2\mu_0} + a_1 e^{-2\sqrt{\mu_0}y} + b_1 e^{2\sqrt{\mu_0}y}, \qquad 0< y<r, \\
F_0(y) = \frac{\mu_1}{2\mu_0} + a e^{-2\sqrt{\mu_0}y} + b e^{2\sqrt{\mu_0}y}, \qquad r<y<L.
\end{align}
The coefficients $a_1, b_1, a$ and $b$ are obtained from 
\begin{align}
\lim_{\epsilon\to 0^+}F_0(\epsilon)=\lim_{\epsilon\to 0^+}F_0(L-\epsilon), \\
\lim_{\epsilon\to 0^+}F_0(r-\epsilon)=\lim_{\epsilon\to 0^+}F_0(r+\epsilon),\\
\lim_{\epsilon\to 0^+}\frac{\mathrm{d} F_0}{\mathrm{d}y}(\epsilon)-\lim_{\epsilon\to 0^+}\frac{\mathrm{d} F_0}{\mathrm{d}y}(L-\epsilon)=2g_{IB},\\
\lim_{\epsilon\to 0^+}\frac{\mathrm{d} F_0}{\mathrm{d}y}(r+\epsilon)-\lim_{\epsilon\to 0^+}\frac{\mathrm{d} F_0}{\mathrm{d}y}(r-\epsilon)=2g_{IB}.
\end{align}
The function $F_1(y)$ reads
\begin{align}
F_1(y) = -\frac{1}{8}\frac{\mu_1^2}{\mu_0^2} -3 a_1 b_1+ (c_1-
\frac{\mu_1 a_1}{\sqrt{\mu_0}}y) e^{-2\sqrt{\mu_0}y} + (d_1+
\frac{\mu_1 b_1}{\sqrt{\mu_0}}y) e^{2\sqrt{\mu_0}y} + \frac{a_1^2}{2} e^{-4\sqrt{\mu_0}y} + \frac{b_1^2}{2} e^{4\sqrt{\mu_0}y}, \qquad 0< y<r, \\
F_1(y) = -\frac{1}{8}\frac{\mu_1^2}{\mu_0^2} -3 a b+ (c-
\frac{\mu_1 a}{\sqrt{\mu_0}}y) e^{-2\sqrt{\mu_0}y} + (d+
\frac{\mu_1 b}{\sqrt{\mu_0}}y) e^{2\sqrt{\mu_0}y} + \frac{a^2}{2} e^{-4\sqrt{\mu_0}y} + \frac{b^2}{2} e^{4\sqrt{\mu_0}y}, \qquad r<y<L,
\end{align}
where $c_1, d_1, c$ and $d$ are determined from the boundary conditions
\begin{align}
\lim_{\epsilon\to 0^+}F_1(\epsilon)=\lim_{\epsilon\to 0^+}F_1(L-\epsilon), \\
\lim_{\epsilon\to 0^+}F_1(r-\epsilon)=\lim_{\epsilon\to 0^+}F_1(r+\epsilon),\\
\lim_{\epsilon\to 0^+}\frac{\mathrm{d} F_1}{\mathrm{d}y}(\epsilon)-\lim_{\epsilon\to 0^+}\frac{\mathrm{d} F_1}{\mathrm{d}y}(L-\epsilon)=2g_{IB}F_0(0),\\
\lim_{\epsilon\to 0^+}\frac{\mathrm{d} F_1}{\mathrm{d}y}(r+\epsilon)-\lim_{\epsilon\to 0^+}\frac{\mathrm{d} F_1}{\mathrm{d}y}(r-\epsilon)=2g_{IB} F_0(r).
\end{align}
The parameter $\mu_1$, which defines the chemical potential, is determined from the normalization condition, i.e., $\int (1+2F_0+F_0^2+2F_1)\mathrm{d}y=L$:
\begin{equation}
\mu_1=\frac{2g_{IB}}{L} -\frac{g_{IB}^2}{2L\sqrt{\mu_0}}+\frac{g_{IB}^2}{2L\sqrt{\mu_0}}e^{-2\sqrt{\mu_0}r}(-1+2\sqrt{\mu_0}r). 
\end{equation}
The corresponding ``two-impurities energy" is 
\begin{equation}
\mu_1 N_{BB}-\frac{\mu_0}{2L}N_{BB}\int_{0}^L(2F_0+5F_0^2+2F_1)\mathrm{d}y=\frac{2g_{IB} N_{BB}}{L} - \frac{g_{IB}^2 N_{BB}}{L\sqrt{\mu_0}}-\frac{g_{IB}^2 e^{-2\sqrt{\mu_0}r}N_{BB}}{L\sqrt{\mu_0}}.
\end{equation}
The first two terms define the energy of two non-correlated impurities~\cite{volosniev2017} up to the $g_{IB}^2$ order. The last term is the Yukawa potential, which defines the effective impurity-impurity interaction, cf.~Ref.~\cite{recati2005}. Note that this potential is accurate only if $g_{IB}\to 0$ defines the smallest energy scale of the problem.
\vspace*{1em}

\begin{figure}[t]
\centering
\includegraphics[width=25em]{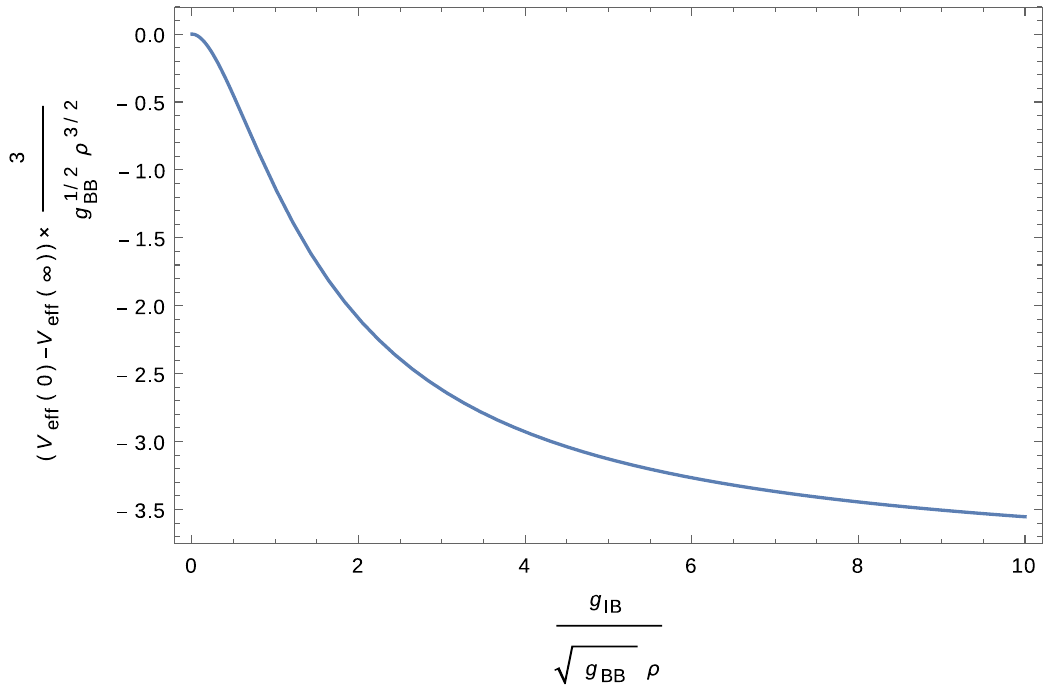}
\caption{The energy release defined in Eq.~(\ref{eq:energy_rel}) as a function of $g_{IB}/(\sqrt{g_{BB}}\rho)$.}
\label{fig:energy_rel}
\end{figure}

{\it Analytical approach.} The Casimir-like force discussed in the paper can be obtained analytically within our model not only for $g_{IB}\to 0$, because the Gross-Pitaevski equation~(\ref{app:eq:gpe}) is solvable on a ring of length $L$ (cf.~Refs.~\cite{volosniev2017,carr2000,malomed2000a}).  The corresponding solution is
\begin{align}
f(y)=\sqrt{\frac{4 K(p_1)^2 p_1}{g_{BB}\delta_1^2r^2(N_{BB}-1)}}\mathrm{sn}\left(2K(p_1)\left[\frac{y-\frac{r}{2}}{\delta_1 r}+\frac{1}{2}\right]\bigg| p_1\right), \qquad 0\leq y\leq r,
\label{eq:app_f_first}\\
f(y)=\sqrt{\frac{4 K(p_2)^2 p_2}{g_{BB}\delta_2^2(L-r)^2(N_{BB}-1)}}\mathrm{sn}\left(2K(p_2)\left[\frac{y-\frac{L+r}{2}}{\delta_2 (L-r)}+\frac{1}{2}\right]\bigg| p_2\right), \qquad r< y < L,
\end{align}
where $K(p)$ is the complete elliptic integral of
the first kind, and $\mathrm{sn}(x|p)$ is the Jacobi elliptic function~\cite{abram1982}. The parameters $p_1, p_2, \delta_1$ and $\delta_2$ are determined from  normalization and the boundary conditions, i.e.,
\begin{align}
\int_{0}^Lf(y)^2\mathrm{d}y&=1, \\
\lim_{\epsilon\to 0^+}f(r-\epsilon)&=\lim_{\epsilon\to 0^+} f(r+\epsilon), \\
2g_{IB}f(r)&=\lim_{\epsilon\to 0^+} \frac{\mathrm{d}f}{\mathrm{d}y}\bigg|_{r+\epsilon}-\lim_{\epsilon\to 0^+}\frac{\mathrm{d}f}{\mathrm{d}y}\bigg|_{r-\epsilon}, 
\end{align}
here $0^+$ means that we deal with one-sided limits. In addition, we have the condition from the chemical potential 
\begin{equation}
\mu = \frac{2 K(p_1)^2(1+p_1)}{\delta_1^2 r^2}=\frac{2 K(p_2)^2(1+p_2)}{\delta_2^2 (L-r)^2}.
\label{eq:app_chem_pot}
\end{equation}
Note that these quantities are available in the literature for a Bose gas with one impurity particle~\cite{volosniev2017}, which can be used to solve the problem with two impurities if $r=0$ and $r\to \infty$. This provides us with the energy release $V_{eff}(0)-V_{eff}(\infty)$ for two static impurities. Indeed, the ``one-impurity" energy is~\cite{volosniev2017}
\begin{equation}
\varepsilon(g_{IB})-\varepsilon(g_{IB}=0)=\frac{\rho^2}{3}\sqrt{\frac{g_{BB}}{\rho}}\left[4+\left(-4+\mathrm{sech}^2(d)\right)\tanh(d)\right], \; \mathrm{where}\;\; d=\frac{1}{2}\mathrm{asinh}\left(\frac{2\rho}{g_{IB}}\sqrt{g_{BB}}\right),
\end{equation}
here $\mathrm{sech}(x),\tanh(x)$ and $\mathrm{asinh}(x)$ are standard hyperbolic
functions. The energy release is then
\begin{equation}
V_{eff}(0)-V_{eff}(\infty)=\frac{\rho^2}{3}\sqrt{\frac{g_{BB}}{\rho}}\left[-4+\left(-4+\mathrm{sech}^2(D)\right)\tanh(D)-\left(-8+2\mathrm{sech}^2(d)\right)\tanh(d)\right],
\label{eq:energy_rel}
\end{equation}
where $D=\frac{1}{2}\mathrm{asinh}\left(\frac{\rho}{g_{IB}}\sqrt{g_{BB}}\right)$.
The function in Eq.~(\ref{eq:energy_rel}) is plotted in Fig.~\ref{fig:energy_rel}, which shows that the energy release is always negative, thus, two static impurities minimize the energy when they cluster; in the limit $g_{IB}\to\infty$ the value of $V_{eff}(0)-V_{eff}(\infty)$ is determined by the activation energy of one dark soliton, which is $-4\rho^{3/2}\sqrt{g_{BB}}/3$ (see below). 

{\it Impenetrable impurity.} In general it is a difficult task to find the form of the potential. Here we show the derivations for impenetrable impurities, i.e., $g_{IB}\to\infty$, assuming that $r$ is much larger than all relevant [physical] length scales. Note that in the thermodynamic limit $p_i=1-m_i$ ($0< m_i\ll 1$) for $i=1,2$ and from Eq.~(\ref{eq:app_chem_pot}) we obtain
\begin{equation}
m_1\simeq 16e^{-r\delta_1\sqrt{\mu}}, \qquad m_2\simeq 16e^{-(L-r)\delta_2\sqrt{\mu}}.
\end{equation}
For impenetrable impurities, $\delta_1=\delta_2=1$ and the solution of the Gross-Pitaevski equation has a much simpler form
\begin{align}
f(y)=\sqrt{\frac{4 K(p_1)^2 p_1}{g_{BB}r^2(N_{BB}-1)}}\mathrm{sn}\left(2K(p_1)\frac{y}{r}\bigg| p_1\right), \qquad 0\leq y\leq r,\\
f(y)=\sqrt{\frac{4 K(p_2)^2 p_2}{g_{BB}(L-r)^2(N_{BB}-1)}}\mathrm{sn}\left(2K(p_2)\frac{y-r}{L-r}\bigg| p_2\right), \qquad r\leq y \leq L.
\end{align}
 The parameters $p_1$ and $p_2$ are related to the chemical potential $\mu = 2 K(p_1)^2(1+p_1)/r^2=2 K(p_2)^2(1+p_2)/(L-r)^2$ through the normalization condition
\begin{equation}
\int_{0}^Lf(y)^2\mathrm{d}y=1\to \frac{4 K(p_1)(K(p_1)-E(p_1))}{g_{BB}r(N_{BB}-1)}+\frac{4 K(p_2)(K(p_2)-E(p_2))}{g_{BB}(L-r)(N_{BB}-1)}=1,
\label{eq:app_norm}
\end{equation}
where $E(p)$ is the complete elliptic function of the second kind~\cite{abram1982}. {\color{blue}Now we insert $p_i=1-m_i$ ($0< m_i\ll 1$) for $i=1,2$ and  rewrite Eq.~(\ref{eq:app_norm}) as
\begin{align}
\frac{8\ln 2(-1+\ln 4)+\ln m_1 (2-8\ln 2+\ln m_1)+(1-4\ln 2+\ln m_1)m_1}{g_{BB}r(N_{BB}-1)} \nonumber\\ + \frac{8\ln 2(-1+\ln 4)+\ln m_2 (2-8\ln 2+\ln m_2)+(1-4\ln 2+\ln m_2)m_2}{g_{BB}(L-r)(N_{BB}-1)} + \mathcal{O}\left(\frac{m_1^2}{r}\right) + \mathcal{O}\left(\frac{m_2^2}{L-r}\right)=1.
\label{eq:condition_m1m2}
\end{align}
The parameters $m_i$ are linked to the chemical potential
($m_1=16e^{-r\sqrt{\mu}}, \qquad m_2=16e^{-(L-r)\sqrt{\mu}}$).
We use these expressions in Eq.~(\ref{eq:condition_m1m2}) to obtain the equation for the chemical potential
\begin{equation}
\frac{-\sqrt{\mu}r(2-\sqrt{\mu}r)+16(1-\sqrt{\mu}r)e^{-\sqrt{\mu}r}}{g_{BB}r(N_{BB}-1)}+\frac{-\sqrt{\mu}(L-r)(2-\sqrt{\mu}(L-r))+16(1-\sqrt{\mu}(L-r))e^{-\sqrt{\mu}(L-r)}}{g_{BB}(L-r)(N_{BB}-1)}=1,
\end{equation}
from which we obtain 
\begin{equation}
\mu\simeq \frac{g_{BB}(N_{BB}-1)}{L}+\frac{4\sqrt{\gamma}\rho^2}{N_{BB}}+\frac{16\rho^2e^{-\sqrt{\gamma}\rho r}}{N_{BB}}\left[\sqrt{\gamma}-\frac{1}{\rho r }\right].
\label{eq:app_chemical}
\end{equation}
The comparison of this result with the chemical potential of the $1+N_B$ system~\cite{volosniev2017} reveals that the last term in Eq.~(\ref{eq:app_chemical}) describes the impurity-impurity correlations. To calculate the effective potential between impurities we calculate the ``two-impurities energy"
\begin{equation}
\left(\mu-\frac{g_{BB} (N_{BB}-1)}{2 L}\right)N_{BB}-\frac{g_{BB}N_{BB}(N_{BB}-1)}{2}\int_{0}^L f(y)^4 \mathrm{d}y=\frac{\rho^2}{\gamma}\left( 8\frac{\gamma^{3/2}}{3}+\frac{16\gamma(\sqrt{\gamma}\rho r-2)e^{-\sqrt{\gamma}\rho r}}{\rho r}\right).
\end{equation}
The quantity $4\frac{\gamma^{1/2}}{3}\rho^2$ is the energy of a dark soliton, which is also the energy of a static impenetrable impurity in a Bose gas~\cite{gaudin1971}. Therefore, we see that at $r\gg 1/(\gamma \rho)$ the impurity-impurity interaction is described by the repulsive $16\rho^2 \sqrt{\gamma} e^{-\sqrt{\gamma}\rho r}$ potential. }

The highlighted discussion from the original submission contains an error that was clarified in the Erratum to the Letter. Below, we present the discussion from the Erratum.  
  
According to the Supplemental Material (SM), the chemical potential for large systems ($L\to\infty$) can be written as
\begin{equation}
\mu\simeq \frac{4K(p_2)^2}{(L-r)^2},
\end{equation}
where $r$ is the distance between impurities. Using this expression, we derive:
\begin{equation}
\frac{\mu}{\mu_0}=1 +\frac{1}{\sqrt{\mu_0}L}\left[2+r\sqrt{\mu_0}-\frac{4K(p_1)(K(p_1)-E(p_1))}{\sqrt{\mu_0}r}\right],
\label{eq:mu_general}
\end{equation}
where $\mu_0$ is the chemical potential without impurities.

To calculate $p_1$, we notice that $r=\sqrt{2(1+p_1)/\mu}K(p_1)$. Therefore, for $r\to\infty$, we must have $p_1\to 1$, which leads to
\begin{equation}
r\sqrt{\mu}\simeq \ln\left(\frac{16}{1-p_1}\right) \to p_1\simeq 1-16e^{-\sqrt{\mu}r}.
\label{eq:expansion}
\end{equation}
If we use this expansion directly in Eq.~(\ref{eq:mu_general}), then we obtain the chemical potential presented in the SM
\begin{equation}
\frac{\mu}{\mu_0}=1 +\frac{1}{\sqrt{\mu_0}L}\left[4+16e^{-\sqrt{\mu_0}r}\left(1-\frac{1}{\sqrt{\mu_0}r}\right)\right].
\label{eq:mu_large_distances_wrong}
\end{equation} 
However, it turns out that this result is not correct as higher-order terms in the expansion~(\ref{eq:expansion}) modify it. 

The simplest way to derive the correct expression for the chemical potential in the limit $r\to\infty$ is to re-write Eq.~(\ref{eq:mu_general}) as
\begin{equation}
\frac{\mu}{\mu_0}=1+\frac{1}{\sqrt{\mu_0}L}\left[2+\sqrt{2(1+p_1)}K(p_1)-\frac{4(K(p_1)-E(p_1))}{\sqrt{2(1+p_1)}}\right],
\label{eq:mu_general_latest}
\end{equation} 
where $r$ has been excluded using $r\simeq \sqrt{2(1+p_1)/\mu_0}K(p_1)$. Equation~(\ref{eq:mu_general_latest}) depends on a single variable ($p_1$), allowing us to avoid any self-consistency issues when using the expansion~(\ref{eq:expansion}). We derive
\begin{equation}
\frac{\mu}{\mu_0}\simeq 1 +\frac{1}{\sqrt{\mu_0}L}\left[4+16\sqrt{\mu_0} r e^{-2 \sqrt{\mu_0}r}-24e^{-2 \sqrt{\mu_0}r}\right].
\end{equation}
The procedure outlined above allows us to compute the tail of the induced impurity-impurity interaction, which is $-16\sqrt{\mu_0}\rho e^{-2\sqrt{\mu_0}r}$ ($\rho$ is the density of the Bose gas without the impurities). Therefore, the impurity-impurity potential is attractive for particles placed far from each other.


\begin{thebibliography}{99}
\bibitem{footnote0} If the impurity particles have some non-trivial structure (e.g., if they are molecules), other models may be employed; see~\cite{schmidt2017}.
\bibitem{footnote4} Note that this equation can be also used to study low-lying excited states associated with impurity dynamics.
\bibitem{fermions} In practice, we obtain the reference energy by solving the $2+N$ problem where one impurity is non-interacting and then use the lowest states to compose the wave function for bosonic and fermionic impurities.
\bibitem{footnote_bind_energy} Assuming that the Yukawa potential represents well the induced interaction also for mobile impurities, we derive that the binding energy is~\cite{landau1977} $\simeq m_B(\int V\mathrm{d}x)^2/4\hbar^2= m_B g_{IB}^4/(4 g_{BB}^2\hbar^2)$ for $m_I=m_B$. To estimate this number in cold-atom systems we set $g_{IB}=g_{BB}$, take $g_{BB}=3\hbar^2/(m_B \mu \mathrm{m})$ and $m_B$ for $^{87}\mathrm{Rb}$ as in Ref.~\cite{catani2012}. These assumptions imply that this energy scale is $\sim 15 n\mathrm{K}\times k_{B}$, meaning that ring traps with temperatures below this scale can be a platform for studying induced impurity-impurity interactions.
\end{thebibliography}
\end{document}